
\input harvmac
\noblackbox

\def \FM {$F$-model\  }
\def \KM {$K$-model\  }

\def \eq#1 {\eqno {(#1)}}

\def\bl{\left}
\def\br{\right}

\def \ra {\rightarrow}
\def\k{\kappa}
\def\r{\rho}
\def\a{\alpha}
\def\b{\beta}

\def\g{\gamma}

\def\d{\delta}

\def\e{\epsilon}

\def\p{\phi}

\def\th{\theta}

\def\m{\mu}
\def\n{\nu}

\def\l{\lambda}

\def\s{\sigma}

\def \sm {$\s$-model\ }

\def \bd {\bar \del}
\def \bh { {\bar h} }

\def \A { {\bar A} }

\def \ha {{1\over 2}}

\def \ov {\over}

\def\const{{\rm const}}
\def \p {\phi}
\def \vp {\varphi}

\def\bl{\bigl}
\def\br{\bigr}

\def \sms {$\s$-models\ }

\def \bd {\bar \del}

\def \tt {{\tilde \t}}

\def \ra {\rightarrow}

\def \na {\nabla }

\def \a {\alpha}
\def \b {\beta}

\def \Tr {{\ \rm Tr \ }}

\def \ln {{\rm \ ln \  }}
\def \det {{\ \rm det \ }}
\def \ch {{\rm cosh \ }}
\def \th {{\rm tanh  }}
\def \l {\lambda}
\def \p {\phi}

\def \m {\mu }
\def \n {\nu}
\def \ep {\epsilon}
\def\g {\gamma}
\def \r {\rho}
\def \k {\kappa }
\def \d {\delta}
\def \o {\omega}
\def \s {\sigma}
\def \t {\theta}

\def \fourth {{\textstyle{1\over 4}}}
\def \six {{\textstyle{1\over 6}}}

\def \e#1 {{{\rm e}^{#1}}}
\def \const {{\rm const }}

\def \eq#1 {\eqno {(#1)}}
\def \sm {$\s$-model\ }

\def \bd  {{ \bar \del }}

\def \bd  { \bar \del }

\def \A { \bar A}


\def \o {\omega}

\def \p {\phi}
\def \ep {\epsilon}
\def \s {\sigma}

\def \r {\rho}
\def \d {\delta}
\def \l {\lambda}
\def \m {\mu}

\def \g {\gamma}
\def \n {\nu}

\def \fourth {{1\over 4}}

\def \e#1 {{{\rm e}^{#1}}}
\def \const {{\rm const }}\def \vp {\varphi}

\def \m {\mu}  

\def \ep {\epsilon}

\def \ra {\rightarrow}

\def \const {{\rm const} }

\def \eq#1 {\eqno{(#1)}}
\def \e {\rm e}
\def \ra {\rightarrow }
\def \e#1 {{\rm e}^{#1}}

\def \ch {\ {\rm cosh} \ }
\def \th {\ {\rm tanh} \ }
\def \ln { {\rm ln } }
\def \sin {\ {\rm sin} \ }
\def \cos { {\rm cos}  }

\def \l {\lambda}
\def \p {\phi}
\def \vp {\varphi}
\def  \g {\gamma}
\def \o {\omega}
\def \r {\rho}

\def\({\left (}
\def\){\right )}
\def\[{\left [}
\def\]{\right ]}

\def \tt {\theta}

\def \ttt {{\tilde \theta }}
\def\bd {{\bar \del}}\def \ra {\rightarrow}
\def \vp {\varphi}

\def \eq#1 {\eqno{(#1)}}

\def \a {\alpha}

\def \b {\beta}
\def \k {\kappa}

\def \o {\omega}

\def \p {\phi}
\def \ep {\epsilon}
\def \s {\sigma}
\def \r {\rho}
\def \d {\delta}
\def \l {\lambda}
\def \m {\mu}
\def \g {\gamma}
\def \n {\nu}

\def \fourth {{1\over 4}}

\def \e#1 {{\rm e}^{#1}}
\def \const {{\rm const }}

\def \vp {\varphi}

\def \ha { { 1\over 2 }}

\def \ov {\over}

\def \sm  { sigma model\ }
\def \at { antisymmetric tensor \ }

\def\np {  Nucl. Phys. }
\def \pl { Phys. Lett. }
\def \mpl { Mod. Phys. Lett. }
\def \prl { Phys. Rev. Lett. }
\def \pr  { Phys. Rev. }

\def \cmp { Commun. Math. Phys. }
\def \ijmp { Int. J. Mod. Phys. }
\def \cqg {Class. Quant. Grav.}
\lref \tss { A.A. Tseytlin, \pl  B317(1993)559. }
\lref \kltss {  C. Klim\v c\'\i k  and A.A. Tseytlin, \pl B323(1994)305;
hep-th/9311012.}

\lref \jack {I. Jack, D.R.T. Jones and N.  Mohammedi, \np B332(1990)333.}
\lref \calg { C.G. Callan and Z. Gan, \np B272(1986)647. }
\lref \ttss {A.A. Tseytlin, \pl B178(1986)34. }
\lref \fri  {D.H. Friedan, \prl 45(1980)1057; Ann. Phys. (NY) 163(1985)318. }
\lref \tsss { A.A. Tseytlin, \ijmp A4(1989)1257. }
\lref \osb { H. Osborn, Ann. Phys. (NY) 200(1990)1.}
\lref \jackk {I. Jack, D.R.T. Jones and D.A. Ross, \np B307(1988)130.}
\lref \jackkk {I. Jack, D.R.T. Jones and N. Mohammedi, \np B332(1990)333.}
\lref \shts  {A.S. Schwarz and A.A. Tseytlin, \np B399(1993)691.}

\lref \berg {E.A. Bergshoeff, R. Kallosh and T. Ortin, \pr D47(1993)5444. }

 \lref \kumar {A. Kumar,
\pl B293(1992)49; D. Gershon, preprint TAUP-2005-92.}
\lref  \hussen {  S. Hussan and A. Sen,  \np B405(1993)143. }
\lref \kiri {E. Kiritsis, \np B405(1993)109. }
\lref \alv { E. Alvarez, L. Alvarez-Gaum\'e, J. Barb\'on and Y. Lozano,
preprint CERN-TH.6991/93.}
\lref \sfts { K. Sfetsos, to appear. }
\lref \nap { M. Henningson and C. Nappi,  \pr D48(1993)861.  }
\lref \Ve    { K.M. Meissner and G. Veneziano, \jnl \pl,  B267, 33, 1991;
A. Sen,  \jnl \pl,   B271,  295, 1991.}

\lref \givkir {A. Giveon and E. Kiritsis,  preprint CERN-TH.6816/93,
RIP-149-93. }\lref  \rocver { M. Ro\v cek and E. Verlinde, \jnl \np, B373, 630,
1992.}

\lref \brink{ H.W. Brinkmann, {\it Math. Ann.} {\bf 94} (1925) 119.}
\lref \guv {R. Guven, Phys. Lett. B191(1987)275.}

\lref \amkl { D. Amati and C. Klim\v c\'\i k, \pl B219(1989)443.}

\lref \hor { G. Horowitz and A.R. Steif, Phys.Rev.Lett. 64(1990)260;
Phys.Rev. D42(1990)1950.}

\lref \horr {G. Horowitz, in: {\it Proceedings
of  Strings '90},
College Station, Texas, March 1990 (World Scientific,1991).}

\lref \rudd { R.E. Rudd, \np B352(1991)489 .}

\lref \duval { C.Duval, G.W. Gibbons and P.A. Horv\'athy, \pr D43(1991)3907 ;
C.Duval, G.W. Gibbons, P.A. Horv\'athy and M.J. Perry, unpublished (1991);
C. Duval, Z. Horvath and P.A. Horv\'athy, \pl B313(1993)10; Marseille
preprints
 (1993). }

\lref \tsnuu { A.A. Tseytlin,  \jnl \pr,  D47, 3421, 1993.}
\lref \tsnull { A.A. Tseytlin, \np B390(1993)153.}

\lref \dunu { G. Horowitz and A.A. Steif, \pl B250(1990)49.}
\lref \pol{ E. Smith and J. Polchinski, \pl
B263(1991)59.}

\lref \hhtt{ J. Horne, G. Horowitz and A. Steif, \prl 68(1992)568.
}
\lref \love {C. Lovelace, \pl B135(1984)75; \np B273(1986)413.}
\lref \FT {E.S. Fradkin and A.A. Tseytlin, \pl B158(1985)316; \np B261(1985)1.}

 \lref \busc { T.H. Buscher, \pl B194(1987)59 ; \pl B201(1988)466.}
\lref \pan  { J. Panvel, \pl B284(1992)50. }

\lref \mye {  R. Myers, \pl B199(1987)371;
    I. Antoniadis, C. Bachas, J. Ellis, D. Nanopoulos,
\pl B211(1988)393;
 \np B328(1989)115. }
\lref \givpas { A. Giveon and A. Pasquinucci, \jnl \pl,  B294, 162, 1992.  }
\lref \GK {A. Giveon and E. Kiritsis,  preprint CERN-TH.6816/93,
RIP-149-93.}
\lref \givroc {A. Giveon and M. Ro\v{c}ek, Nucl. Phys. B380(1992)128.}
\lref \giv {  A. Giveon, \jnl \mpl, A6, 2843, 1991. }

\lref \bars {I. Bars, preprint USC-91-HEP-B3;
 E. Kiritsis, \mpl A6(1991)2871. }

\lref \GRV {A. Giveon, E. Rabinovici and G. Veneziano, Nucl. Phys.
B322(1989)167;
A. Shapere and F. Wilczek, \np B320(1989)669.}
\lref \GMR {A. Giveon, N. Malkin and E. Rabinovici, Phys. Lett. B238(1990)57.}

\lref \VV    {  G. Veneziano, \jnl \pl,   B265,  287, 1991.}

\lref  \horne { J.H. Horne and G.T. Horowitz, \np B368(1992)444.}

\lref \koki { K. Kounnas and  E. Kiritsis, preprint CERN-TH.7059/93;
hep-th/9310202. }

\lref \tsmpl {A.A. Tseytlin, \mpl A6(1991)1721.}

\lref \tsbh {A.A. Tseytlin,  preprint CERN-TH.6970/93; hep-th/9308042.}

\lref \nsw {  K.S. Narain, M.H. Sarmadi and E. Witten, \np B279(1987)369. }

\lref \STT {K. Sfetsos and A.A. Tseytlin, preprint CERN-TH.6969/93,
hep-th/9310159.}
\lref \givkir {A. Giveon and E. Kiritsis,  preprint CERN-TH.6816/93;
hep-th/9303016. }

\lref \cec { S. Cecotti, S. Ferrara and L. Girardello, \np B308(1988)436. }

\lref \tsdu { A.A. Tseytlin, \pl B242(1990)163; \np B350(1991)395.  }


\lref \wi { E. Witten, unpublished (1991). }
\lref \kik {  K.~Kikkawa and M.~Yamanaka,  \jnl \pl, B149, 357, 1984;
N.~Sakai and I.~Senda, {{\it  Progr. Theor. Phys.}}
{{\bf 75}} (1986) 692;
M.~B.~Green,
 J.~H.~Schwarz and L.~Brink, {{\it Nucl.Phys}}.  {{\bf B198}} (1982) 474. }
\lref\nair{ V.~Nair, A.~Shapere, A.~Strominger and F.~Wilczek,
{{\it Nucl. Phys}}. {{\bf B287}} (1987) 402.
}

\lref \vaf {R. Brandenberger and C. Vafa, \jnl \np, B316,  391, 1988; A.
Tseytlin and C. Vafa, \jnl \np,  B372, 443, 1992.}
\lref \por{ A. Giveon, M. Porrati and E. Rabinovici, preprint RI-1-94,
hep-th/9401139. }
\lref \hht{G. Horowitz and A. Steif, \jnl \pl,  B258, 91, 1991.}

\lref \klits{  C. Klim\v c\'\i k  and A.A. Tseytlin, unpublished (1994). }


\baselineskip8pt
\Title{\vbox
{\baselineskip 6pt{\hbox{  }}{\hbox
{Imperial/TP/93-94/46}}{\hbox{hep-th/9407099}} } }
{\vbox{\centerline {  Exact  string solutions and  duality }
 }}
\vskip 37 true pt
\centerline{   A.A. Tseytlin\footnote{$^{\star}$}{\baselineskip8pt
e-mail address: tseytlin@ic.ac.uk}\footnote{$^{\dagger}$}{\baselineskip5pt
On leave  from Lebedev  Physics
Institute, Moscow, Russia.} }

\smallskip\smallskip
\centerline {\it  Theoretical Physics Group, Blackett Laboratory}

\centerline {\it  Imperial College,  London SW7 2BZ, U.K. }
\medskip
\centerline {\bf Abstract}
\smallskip
\baselineskip6pt
\noindent
We  review known exact  classical solutions in  (bosonic) string theory.
The  main classes of the solutions are
`cosets' (gauged WZW models), `plane wave'-type backgrounds (admitting a
covariantly constant null Killing vector) and  `$F$-models' (backgrounds with
two null Killing vectors generalising the `fundamental string' solution).
The recently constructed  $D=4$ solutions with  Minkowski signature
are given explicitly.
We consider various  relations  between  these solutions
and, in particular,
 discuss some aspects of the duality symmetry.

\bigskip\bigskip\bigskip

\centerline{ To appear in: {\it Proceedings of the 2nd Journ\'ee Cosmologie},}
\centerline{\it Observatoire de Paris, June 2-4, 1994 (World Scientific,
Singapore)}

\Date {July 1994}

\noblackbox
\baselineskip 16pt plus 2pt minus 2pt
\noblackbox

\vfill\eject


\def \lr { \lref}

\gdef \jnl#1, #2, #3, 1#4#5#6{ {\sl #1~}{\bf #2} (1#4#5#6) #3}

\lr \mans { P. Mansfield and J. Miramontes, \jnl \pl, B199, 224, 1988;
A. Tseytlin, \jnl \pl, B208, 228, 1988; \jnl \pl, B223, 165, 1989.}

\lr \kalmor{R. Kallosh and A. Morozov,  \jnl \ijmp,  A3, 1943, 1988.}

\lr \ghrw{J. Gauntlett, J. Harvey, M. Robinson, and D. Waldram,
\jnl \np, B411, 461, 1994.}
\lr \garf{D. Garfinkle, \jnl \pr, D46, 4286, 1992.}

\lr \onofri { V. Fateev, E. Onofri and Al. Zamolodchikov, \jnl \np, B406,
 521, 1993.}

\lref \tspl {A. Tseytlin, \jnl \pl, B317, 559, 1993.}
\lref \tssfet { K. Sfetsos and A.  Tseytlin, \jnl  \pr, D49, 2933, 1994.}
\lref \klts {C. Klim\v c\'\i k  and A. Tseytlin, ``Exact four dimensional
string solutions and Toda-like sigma models from null-gauged
WZNW models",  preprint
 Imperial/TP/93-94/17, hep-th/9402120.}

\lr \sfexac {K. Sfetsos,  \jnl \np, B389, 424,  1993.}

\lr \tsmac{A. Tseytlin, \jnl \pl,  B251, 530, 1990.}

\lr \cakh{C. Callan and R. Khuri, \jnl \pl, B261, 363, 1991;
R. Khuri, \jnl \np, B403, 335, 1993.}
\lr \dgt{M. Duff, G. Gibbons and P. Townsend, ``Macroscopic superstrings
as interpolating solitons", DAMTP/R-93/5, hep-th/9405124.}

\lref \ger {A. Gerasimov, A. Morozov, M. Olshanetsky, A. Marshakov and S.
Shatashvili, \jnl \ijmp,
A5, 2495,  1990. }

\lr \hutow{C. Hull and P. Townsend, \jnl \np, B274, 349, 1986.}
\lr \mukh {S. Mukhi,    \jnl \pl,  B162, 345, 1985;
S. De Alwis, \jnl \pl, B164, 67, 1985. }

\lr \napwietc { C. Nappi and E. Witten, \jnl \prl, 71, 3751, 1993;
E. Kiritsis and C. Kounnas, \jnl \pl, B320, 264, 1994; D.  Olive,
 E. Rabinovici and A. Schwimmer,  \jnl \pl, B321, 361, 1994;
 K. Sfetsos,  \jnl \pl, B324, 335, 1994;
  ``Gauged WZW models and Non-abelian Duality'',
THU-94/01, hep-th/9402031;
I. Antoniadis and N. Obers, ``Plane Gravitational Waves in String
Theory'', CPTH-A299.0494, hep-th/9403191;
 J.M. Figueroa-O'Farrill and S. Stanciu, QMW-PH-94-2, hep-th/9402035; N.
Mohammedi, BONN-HE-93-51, hep-th/9312182;
A.  Kehagias and P.  Meessen, \jnl \pl,  B320, 77, 1984;
A.  Kehagias, NTUA-45/94, hep-th/9406136.}

\lr \sfeets{
K. Sfetsos and A. Tseytlin,  ``Four Dimensional Plane Wave String Solutions
with Coset CFT Description", preprint THU-94/08, hep-th/9404063. }

\lr \scherk { J. Scherk, \jnl \np, B31,  222, 1971;
   J. Scherk and J. Schwarz, \jnl \np, B81, 118,
1974;  T. Yoneya, \jnl {\it Progr. Theor. Phys.}, 51, 1907, 1974.}
 \lr \lov  {C. Lovelace,  \jnl \np,  B273, 413,  1986.}
\lr \call{C. Callan, D. Friedan, E. Martinec and  M. Perry, \jnl \np, B262,
593, 1985.}
\lr \frts {E.  Fradkin  and A. Tseytlin, \jnl \pl, B158, 316, 1985;
\jnl \np, B261, 1, 1985.}
\lr \tsred  {A. Tseytlin, \jnl  \pl, B176, 92, 1986; \jnl  \np, B276, 391,
 1986.}
\lr \grwi   { D. Gross and E. Witten, \jnl \np, B277, 1, 1986.}

\lr \gps {S.  Giddings, J. Polchinski and A. Strominger, \jnl  \pr,  D48,
 5784, 1993. }

\lr \tsppl  {A. Tseytlin, \jnl   \pl,  B208, 221, 1988.}
\lr\rabi  {S. Elitzur, A. Forge and E. Rabinovici, \jnl \np, B359, 581, 1991;
 G. Mandal, A. Sengupta and S. Wadia, \jnl \mpl,  A6, 1685, 1991. }
 \lr \witt{ E. Witten, \jnl \pr, D44, 314, 1991. }
 \lr \dvv { R. Dijkgraaf, H. Verlinde and E. Verlinde, \jnl \np, B371,
269, 1992.}
\lr \hoho { J. Horne and G.  Horowitz, \jnl \np, B368, 444, 1992. }
\lr \horwel{G. Horowitz and D. Welch, \jnl \prl, 71, 328, 1993;
N. Kaloper,  \jnl \pr,  D48, 2598, 1993. }
\lr \host{ G. Horowitz and A. Steif,  \jnl \prl, 64, 260, 1990; \jnl \pr,
D42, 1950, 1990;  G. Horowitz, in: {\it
 Strings '90}, (eds. R Arnowitt et. al.)
 World Scientific (1991).}
\lr \busch {T.  Buscher, \jnl \pl, B194, 59, 1987; \jnl \pl,
 B201, 466, 1988.}
\lr \kallosh {E. Bergshoeff, I. Entrop, and R. Kallosh, ``Exact Duality in
String Effective Action", SU-ITP-93-37; hep-th/9401025.}

\lr \tsmpl {A. Tseytlin, \jnl  \mpl, A6, 1721, 1991.}
\lr \vene { }
\lr \kltspl { C. Klim\v c\'\i k and A. Tseytlin, \jnl \pl, B323, 305, 1994.}
\lr \shwts { A. Schwarz and A. Tseytlin, \jnl \np, B399, 691, 1993.}
\lr \callnts { C. Callan and Z. Gan, \jnl  \np, B272, 647, 1986;  A. Tseytlin,
\jnl \pl,
B178, 34, 1986.}

\lr \guv { R. G\"uven, \jnl \pl, B191, 275, 1987;
 D. Amati and C. Klim\v c\'\i k,
\jnl \pl, B219, 443, 1989;
 R. Rudd, \jnl \np, B352, 489, 1991.}
\lr \desa{ H. de Vega and N. Sanchez, \jnl
\pr, D45, 2783, 1992; \jnl \cqg, 10, 2007, 1993.}
\lr \desas{ H. de Vega and N. Sanchez, \jnl
\pl, B244,  215, 1990.}
\lref \tsnul { A. Tseytlin, \jnl \np, B390, 153, 1993.}

\lref \dunu { G. Horowitz and A. Steif, \pl B250 (1990) 49;
 E. Smith and J. Polchinski, \pl B263 (1991) 59. }

\lr \gauged {I. Bars and K. Sfetsos, \jnl  \mpl, A7, 1091, 1992;
 P. Ginsparg and F. Quevedo, \jnl \np, B385, 527, 1992. }
\lr \bsfet {I. Bars and K. Sfetsos, \jnl \pr, D46, 4510, 1992; \jnl \pr,
 D48, 844, 1993. }
\lr \tsnp{ A. Tseytlin, \jnl \np, B399, 601, 1993;  \jnl \np, B411, 509, 1994.}
\lr \gibb{A. Dabholkar, G. Gibbons, J. Harvey, and F. Ruiz, \jnl \np, B340,
33, 1990.}
\lr \hhs{J. Horne, G. Horowitz, and A. Steif, \jnl \prl, 68, 568, 1992;
G. Horowitz, in:  {\it  Proc. of the 1992 Trieste Spring School on String
theory  and Quantum Gravity},
preprint UCSBTH-92-32, hep-th/9210119.}

\lr \polypolchnats {  }
\lr \jack {I. Jack, D.  Jones and J. Panvel, \jnl \np, B393, 95, 1993.}
\lr \mettstwo {R.  Metsaev and A. Tseytlin, \jnl \pl, B185, 52, 1987.}
\lr \banks {T. Banks, M. Dine, H. Dijkstra and W. Fischler, \jnl \pl, B212,
45, 1988.}
\lr \horstr{G. Horowitz and A. Strominger, \jnl \np, B360, 197, 1991.}

\lr \givkir {A.  Giveon and E. Kiritsis, \jnl \np, B411, 487, 1994.  }

\lr \jac{I. Jack and D. Jones, \jnl \pl, B200, 453, 1988.}
\lr \metts{R. Metsaev and A. Tseytlin, \jnl \np,  B293, 385, 1987.  }
\lr \horv{ P. Horava, \jnl \pl,
B278, 101, 1992.}

\lref \FT {E.S. Fradkin and A.A. Tseytlin, Phys.Lett. B158(1985)316; Nucl.Phys.
B261(1985)1. }
\lref \love {C. Lovelace, \pl B135(1984)75; \np B273(1986)413.}
\lref \sch {J. Scherk and J.H. Schwarz, \np B81(1974)118.}

\lref \por{ A. Giveon, M. Porrati and E. Rabinovici, preprint RI-1-94,
hep-th/9401139. }
\lref \klits{  C. Klim\v c\'\i k  and A.A. Tseytlin, unpublished (1994). }

\lr \horts { G. Horowitz and A.A. Tseytlin, ``On exact solutions and
singularities in string theory", preprint  Imperial/TP/93-94/38;
hep-th/9406067.}

\lr \hort { G. Horowitz and A.A. Tseytlin,  unpublished (1994).}
\lr \hrts { G. Horowitz and A.A. Tseytlin, ``A new class of exact string
solutions",
to appear.}

\lref \gris {  M.T. Grisaru, A. van de Ven and D. Zanon, \jnl \np, B277,  409,
1986. }

\lref \wit {E. Witten, \jnl \cmp,  92, 455, 1984.}
\lref\bra{ E. Braaten, T.L. Curtright and C.K. Zachos, \jnl  \np,  B260,  630,
1985.}

\lr\gepner{  D. Gepner, \jnl \np, B296,  757, 1988. }

\lr \nemsen{D. Nemeschanski and A. Sen, \jnl \pl, B178, 365, 1986.  }
\lr\cres{ K. Bardakci, M. Crescimanno and E. Rabinovici,
{\it Nucl. Phys.}    {\bf B344} (1990) 344.}

\lref \gwzw {      K. Bardakci, E. Rabinovici and
B. S\"aring, \jnl \np, B299, 157, 1988;
 K. Gawedzki and A. Kupiainen,
\jnl \np, B320, 625, 1989;  D. Karabali, Q-Han Park, H.J.
Schnitzer and
Z. Yang, \jnl \pl,  B216,  307,  1989;  D. Karabali and H.J. Schnitzer, \jnl
\np,  B329,  649, 1990.
}
\lref \bha { K. Bardakci and M.B. Halpern, \jnl \pr,  D3, 2493,  1971;
M.B. Halpern, \jnl \pr, D4, 2398, 1971.}

\lref\GKO{P. Goddard, A. Kent and D. Olive, \jnl \pl,  B152,   88,  1985.}

\def \bh{\tilde h}

\lref\kou{C. Kounnas and D. L\"ust, {\it Phys. Lett.}  {\bf B289} (1992) 56.}

\lref\barsf { I. Bars and K. Sfetsos, \jnl \pl,  B277, 269,  1992;
 \jnl \pr,  D46,  4495, 1992; \jnl \pr,  D46,  4510, 1992;
I. Bars, preprint USC-93/HEP-B3,  hep-th/9309042. }

\lref\brsne{ I. Bars and D. Nemeschansky, {\it  Nucl. Phys.}  {\bf B348} (1991)
89.}

\lref \napwit { C. Nappi and E. Witten, \jnl \pl,  B293, 309, 1992.}

\lr\duval  { C. Duval, Z. Horvath and P.A. Horvathy, \jnl \pl,  B313, 10,
1993.}
\lr\bergsh { E. Bergshoeff, R. Kallosh and T. Ort\'in, \jnl \pr,  D47, 5444,
1993.}

\lref \penr {R. Penrose,  `Any space-time has a plane wave as a limit',
in: Differential Geometry and Relativity, A. Cahen and M. Flato eds.,
(Reidel, Dordrecht-Holland), 1976, p. 271. }
\lr\sen{A. Sen, \jnl \np, B388, 457, 1992. }
\lr \garf{D. Garfinkle, \jnl \pr, D46, 4286, 1992.}
\lr \wald {D. Waldram, \jnl \pr, D47, 2528, 1993.}

\newsec{Introduction}

One of the  central  problems in string theory is to
study the space of exact classical solutions.
This may clarify the formal structure of string
 theory and also may be relevant for understanding the implications
of string theory  (assuming there exist regions where string coupling is small
so that perturbative and non-perturbative corrections to string equations of
motion can be ignored).

To be able to discuss  issues of singularities and
short distance structure
one should be interested not just  in  solutions of the leading-order
low-energy  string effective equations (derived under the assumption of small
curvature $|\a'R|\ll  1$) but in the ones  that  are exact
in $\a'$.  In general it is not enough to know just
a set of `massless' background fields $(G_{\m\n},B_{\m\n},\p)$.
Since strings are extended objects  test  strings
`feel'  geometry in a way which is different from point particles.
Different (dual) point-particle geometries may appear to be equivalent
from the string point of view.
In order to  give an adequate description of  a first-quantised string
propagation,
i.e. a `string-geometric' interpretation of a given solution one needs to know
the
corresponding $2d$ conformal field theory (CFT).
The knowledge of CFT includes information about equations for all string modes
in a given background and thus  goes beyond a specification of one  particular
conformal $\s$-model.

The tree-level effective action for the `massless' fields is given
by an infinite powers series in  $\a'$.
While the solutions of the leading-order equations are numerous
and straightforward to find,  only a small subset of them is known to  be exact
to all orders in $\a'$.   Moreover,  only some of these exact solutions have
known CFT interpretation.

Given that the general structure of the effective action is unknown,
to find the exact solutions one needs to use  some indirect methods.
The key  property is that the  string solutions, i.e. the stationary points of
the effective action,  must correspond to conformally invariant sigma models
\refs{\lov,\call}.

There are three possible  approaches that  were   used
to construct exact string solutions:

(i) Start with a particular leading-order solution and show that
$\a'$-corrections
are absent or take some explicitly known form
 due to some special properties of this background.  Such are some
`plane wave'-type  backgrounds, or, more generally, backgrounds  with  a
covariantly constant Killing vector, see e.g.
\refs{\guv,\host,\desa,\tsnul,\horts}.

(ii) Start with a known Lagrangian CFT and represent it as a conformal
$\s$-model.
Essentially the only known example of such construction is based
on gauged WZW models, see e.g.  \refs{\witt,\gauged,\horv, \dvv}.

(iii) Start  directly with  a  leading-order \sm path integral
and prove that there exists such a definition of the \sm couplings (i.e. a
`renormalisation scheme') for which this \sm is  conformal  to all  loop
orders.
This strategy works  \horts\ for so called `$F$-models' introduced in
\refs{\klts, \horts} and their generalisations \hrts.
A particular example of \FM is the `fundamental string' (FS) solution \gibb.

These three classes of exact solutions are not completely independent.
Namely, a subclass of plane waves  can be identified with gauged WZW models
based on
non-semisimple groups \refs{\napwietc, \sfeets}.
Also, a different subclass of backgrounds with a covariantly constant null
Killing vector  (called `$K$-models' in \horts)
is  dual to $F$-models. Finally,  a subclass of $F$-models  can be interpreted
\klits\
as  special  gauged WZW models  (based on maximally non-compact group
and nilpotent subgroup as a gauged one); similar interpretation is true for the
generic $D=3$ \FM \horts.

There exists also a more general class of  exact solutions (with just one null
Killing vector)  which generalises both the  $K$-models and $F$-models  \hrts.

We shall start in  section 2  with a discussion of the structure of the string
effective action stressing the importance of the field redefinition ambiguity
(`scheme dependence').

 In section 3 we shall review some aspects of the duality symmetry
and consider several  $D=3$ examples where the duality is exact to all orders
in $\a'$,
illustrating in which sense  different string modes `feel' different
geometries.

Exact solutions corresponding to gauged WZW models will be considered in
section 4.
We shall  argue that there exists a `leading-order' scheme in which  the
background
fields do not depend on $\a'$. We shall  mention  some  known solutions with
$D=4$.

Solutions with a covariantly constant null Killing vector
will be the topic of section 5. We shall start with the simplest plane wave
backgrounds
(some of which  will be   related to gauged WZW models based on non-semisimple
groups)
and discuss several generalisations,
in particular,  $K$-model dual to the fundamental string solution  and  $D=4$
hybrid $K$-model with the transverse part  represented by the
euclidean $D=2$ background.

Section 6 will be devoted to a  new class of exact solutions described by
$F$-models.
We shall start with the $F$-models with flat transverse part
and  consider  a  subclass of them   (in particular, with  $D=4$) which
corresponds to specific nilpotently gauged WZW models and thus admits
a direct CFT interpretation. We shall also present the    $D=4$
\FM with the $D=2$ euclidean black hole as the transverse part
which generalises the $D=4$ fundamental string solution.

A more general class of exact string solutions which contains both the
$K$-models and $F$-models will be briefly discussed in section 7.

In section 8  we shall make  some concluding remarks, in particular, on the
problem of finding the CFT interpretation of the  exact solutions.

\newsec{Structure
 of the  string effective action }

String effective action contains terms of all orders in $\a'$
and thus depends on an infinite number of parameters.
Some of these parameters are unambiguous, i.e. are determined by the string
$S$-matrix while others  can be changed by local field redefinitions.
The latter  represent the freedom of changing the `scheme'. This `scheme
dependence'  is an
important novel property of the string equations of motion  which needs to be
taken into account in the discussion of exact string solutions: equivalent
exact solutions may look very different in different schemes.
The aim of this section is to discuss the general structure of the
tree-level string theory effective action (EA)
emphasizing a possibility to use the field redefinition freedom
to put higher order $\a'^n$-corrections in the simplest  form.
In particular,  the  EA
 can be chosen in such
a form (a `scheme') that all  $\a'$-corrections  $vanish$ once we specialise to
the case of a $D=2$ background.
In  such a  scheme the $D=2$  metric-dilaton EA is  thus known explicitly,
i.e. is given by the leading-order terms.
There also exists a scheme in which the $D=3$ limit of the EA
has all $\a'$-corrections depending only on the  derivatives of  the dilaton
but not on the curvature or antisymmetric tensor.

\subsec{`Scheme dependence' of the effective action}

Let us first recall a few basic facts about  the string  effective action
\refs{\scherk,\frts}.
Given a tree-level
string S-matrix  (in $D=26$) we can  try to reproduce  its massless sector
by a local covariant field-theory action $S(G,B,\p)$ for the metric,
antisymmetric tensor and dilaton. Subtracting the massless exchanges from the
string scattering amplitudes and expanding the massive ones in powers of $\a'$
gives an infinite series of terms in $S$ of all orders in $\a'$.

 The form of such action is  not  unique:
a class of actions
related by   field redefinitions  which are local,  covariant,
 background-independent, power series in $\a'$ (depending  on dilaton
 only through its derivatives not to mix different orders of string loop
 perturbation theory)  will correspond to the same string $S$-matrix
\refs{\tsred,\grwi}.
Given some representative in a class of equivalent EA's  we refer to
other equivalent actions  as corresponding to different `schemes'.
The reason for this terminology is that the extremality  conditions  for  the
effective action are equivalent
to the conditions of conformal invariance of the
\sm representing string action in a background \refs{\lov,\call}  and the
related ambiguity in the \sm  Weyl anomaly coefficients or
 `$\b$-functions' can be interpreted
as being  a consequence of
 different choices of a  renormalisation scheme \tsred.
This implies that  some coefficients of the $\a'^n$-terms
in the EA will be unambiguous (being fixed by the string $S$-matrix) while many
others  will be `scheme-dependent'.

Though one possible way of determining  the EA is to start with  perturbative
massless string scattering amplitudes on a flat $D=26$ background,
 $S$  must actually be background-independent.
 In particular, its  unambiguous coefficients are
 universal (e.g. they  do not dependent on the dimension $D$). This is implied
by the equivalence between the
effective equations of motion and the string \sm Weyl  invariance
conditions (which are  background-independent).
To  make this  equivalence  precise in any dimension $D$
we need only to  add       to the EA
one  $D$-dependent   (`central charge')
term $\sim \int d^D x \sqrt G \exp (-2\p) \  (D-26) $.
Given such a  background-independent EA \refs{\scherk,\frts,\call}
\eqn\act{ S = \int d^D x \sqrt G \  \e{-2\p}   \ \{ {2(D-26)\ov 3\a' }   -
  \  [ \   R \ + 4 (\del_\m \p )^2 - {1\ov 12} (H_{\m\n\l})^2\  ]
  + O(\a')   \}  \  , }
it  would be useful  to choose a scheme (i.e. the values of
  ambiguous coefficients) in which  $S$ has the
  simplest possible form.
For example, the correspondence with the Weyl  anomaly coefficients of a string
\sm implies that there exists a scheme in which \act\ does not contain other
higher-order dilatonic terms.
This follows  also from the general argument \tsppl\ based on the path integral
representation for the EA \frts\ and was checked directly at the $\a'$-order
\refs{\metts,\jac}  by comparing with   string $S$-matrix (as we shall note
below show
in three dimensions one can do just the opposite, i.e., have only dilaton
terms as higher order corrections).

For general $D\geq 4$ the
 massless sector of string $S$-matrix is non-trivial
so  no simple scheme  is  expected to exist.
To order $\a'$ one  can choose a  `standard' scheme in which the effective
action
has the form \metts
\eqn\actt{  S = \int d^{D}x \sqrt {\mathstrut G} \ {\rm e}^{- 2
\phi} \big\{  {2(D-26)\ov 3\a' }   -
  \  [ \   R \  + 4 \na^2 \p - 4 (\na \p )^2 - {1\ov 12} (H_{\m\n\l})^2\  ]  }
 $$  - {1 \over 4} \a'\  [  \ R_{\m\n\l\k}^2 -\ha R^{\m\n\k\l}H_{\ \m\n}^\r
H_{\k\l\r} $$ $$
+ {1\ov 24} H_{\m\n\l}H^\n_{\ \r\a}H^{\r\s\l}H_\s^{\ \m\a}  - {1\ov 8}
(H_{\m\a\b}H_\n^{\ \a\b})^2\ ] +
O(\a'^3)   \big\} \  .  $$

\subsec{Effective action in  $D\leq 3$}

It is possible   to
arrive at a more definitive conclusion  about a simplest possible scheme
by specialising to the low dimensional cases of
$D=2$ and $D=3$ \horts. More precisely, we would like to find  an  EA
 (defined for  general $D$)  such that  its $\a'^n$-terms take a simple form
in the limit $D\ra 2,3$.

 Given that   \act\ is background-independent (in particular,
its higher-order coefficients do not depend on $D$)
we are free to  take $(G,B)$
in \act\ to correspond to a generic $D=2$ or $D=3$ background.
Since the basic fields $(G,B)$ are second rank tensors, higher order
terms which involve `irreducible' contractions of tensors of rank greater
than two  cannot be altered
by field redefinitions. But
in $D\leq 3$   the Riemann tensor can be expressed in terms
 of the Ricci tensor, and $H_{\m\n\l} = \epsilon_{\m\n\l} H$. Thus
all possible covariant structures in the EA will have the `reducible' form
of products of scalars,  vectors, or at most,  second-rank tensors.

This is a necessary condition for a higher order term to be removed by
a field redefinition, but it is not sufficient. It has been shown \mettstwo\
 that some combinations of a priori ambiguous coefficients in
 the EA are actually redefinition-invariant
(unambiguous) and thus are uniquely  determined by the string $S$-matrix.
In fact, it is easy to show that one cannot  find  a scheme
in which there is no $\a'$-term in the $D=3$ EA \horts.
 However in $D=2$ (where  $H_{\m\n\l}=0$) one can write
 $ (R_{\m\n\k\l})^2 = 2(R_{\m\n})^2$  so the $\a'$ term now vanishes
 when the leading order equations are satisfied. Using the results of
\metts\ one can show that this term
can indeed be removed by a field redefinition.

More generally, suppose we compute the scattering
amplitudes for the dilaton and graviton (in general $D$)
directly in the string frame where the dilaton and  graviton mix
in the propagator.
Since there are  no transverse degrees of freedom for the string in $D=2$,
there are no dynamical degrees of freedom in the $(G,\p)$ system,  and
the  limit $D \ra 2$ of the scattering amplitudes  is  trivial.
That means that  the on-shell limits of unambiguous terms in the EA
  must  vanish identically. Hence
 there exists   a choice of the EA (in generic $D$) such
that    higher order terms in
it  vanish in the $D\ra 2$ limit.\foot{We are not including
massive modes (tachyon, etc) in the effective action since we  are mostly
interested in general  $D$. It is true, however,  that
the tachyon becomes  massless  in $D=2$ and thus  one can define an effective
action which will involve it as well.}

A similar statement is not true in $D=3$ since there is one transverse
degree of freedom for the string which could yield higher order
corrections to the scattering amplitudes and hence to the EA.
 However one can express
these corrections solely in terms of the dilaton.

 Since in $D=2$
since there is a scheme in which all $\a'$ corrections to the EA vanish,
all backgrounds which  solve  the leading-order equations
are in fact   exact solutions.
This conclusion is not  so surprising:  the $D=2$   `black hole'
 background \refs{\rabi,\witt} represents the generic solution
of the leading-order equations, and given that the corresponding CFT is known
($SL(2,R)/U(1) $ coset \witt)
one can find  explicitly \tsppl\  a local covariant background-independent
redefinition from the `CFT scheme'
 \dvv\   (where the background fields are $\a'$-
dependent) to the  `leading-order' scheme.
It also follows that in this scheme the $D=2$  \sm Weyl anomaly
coefficients  just have their leading-order form.

As for $D=3$,  in the scheme where $\a'$-corrections
are proportional to the derivatives of $\p$ the solutions
of the leading-order equations which have $\p=\const$ remain exact
to all orders. It is easy to see that the only  leading-order solution  with
$\p=\const$ in $D=3$
is the constant curvature anti-DeSitter  space with  the parallelising
$H_{\m\n\l}$-torsion corresponding to  the $SL(2,R)$ WZW model or its possible
cosets over discrete subgroups (in particular, the $D=3$  black hole of
\horwel).
For solutions with $\p\not=\const$ the best what we can hope for is to find a
scheme in which a particular
leading-order solution does not receive $\a'$-corrections. This was shown
\tssfet\ to be the case (to $\a'^2$ order) for the
charged black string background  \hoho, i.e.
$SL(2,R)\times R/R $  coset model.
It was also found \horts\  that there  exists a scheme where the leading order
solution
for the $D=3$
$F$-model
is also conformal  to the next order in $\a'$.

Another  implication of the above discussion of the structure of the EA
concerns an exact form of the abelian duality transformations which map string
solutions into solutions (see next section). The
leading-order duality \busch\ is  a  symmetry of the leading-order terms in the
EA \refs{\banks,\tsmpl} and thus  is the  $exact$ symmetry in the simplest
scheme in $D=2$.
In fact, one can check
directly \hort\  that while for a  general $D$ there   does not exist a
scheme
in which the  leading order duality  remains a symmetry  at $\a'$-order
without been modified by the derivative $O(\a')$-term \tsmpl,
such scheme does exist in $D=2$.\foot{The locality of the $\a'$-correction
\tsmpl\
to the duality  (and the existence of several examples  where leading-order
duality relates exact solutions to exact solutions) strongly suggests that it
can also be interpreted as some generalised change of the scheme provided one
somehow relaxes the condition of covariance of the redefinition. }



\newsec{Duality  symmetry}
String theory is a remarkable extension of  General Relativity
which is consistent at the quantum level.
 Being a theory of extended objects
string theory is radically  different  from the  Einstein  theory in its
description of space-time geometry.  Both the  diffeomorphism invariance and
the Einstein action  appear only as  large-distance approximations to the
full string symmetry group  and string field theory action.
The  space-time metric  $G_{\m\n}$    provides  only  a crude
description of the  true  `string geometry'
when probed  by  point-like string  states.

The metric and its curvature  are not  adequate characterisations  of the
string geometry (i.e. they are not  invariants
of the string  symmetry group):  spaces with different curvature
may represent  equivalent string backgrounds.
This  can be demonstrated on the example of particular
string symmetry --  space-time duality symmetry (see, e.g.,
\refs{\kik,\nair,\busch,\VV,\tsmpl,\Ve,\rocver,\por}).

In this section we shall
 present  some  examples of string backgrounds related by duality emphasizing
that even in the simplest   point-like approximation  an  invariant content of
the  string geometry is described  not just by the  metric  but by  a
combination  of the metric,  antisymmetric tensor  and  dilaton with all the
three fields  playing  complementary  roles.
   Particular properties of the metric (e.g. its curvature and causal
structure)  are not, in general,   invariant under duality.  Different string
modes may `see' different  `sides'
of the same string geometry.
 For example,   a flat space  (with some compact dimensions)
viewed as a solution of the string theory can  be equivalent  to a curved space
with a non-trivial dilaton
and/or antisymmetric tensor.  While the  momentum string modes move in the
original flat geometry,
 the winding modes present in the `flat' string spectrum
propagate in a  dual curved  background.   This is certainly very  different
from  what we are used to  in the $field$ theory of gravity -- Einstein theory.

The  examples of backgrounds we shall discuss have null Killing vectors and are
 exact string solutions  (to all orders in $\a'$).

\subsec{General remarks on duality}
The space-time duality is one of the most fundamental
properties of string theory,   a direct consequence of the  $one$-dimensional
extended nature of strings.
In $two$ dimensions a scalar field is dual in two-dimensional sense to another
scalar field ($\del_m x = \epsilon_{mn} \del^n \tilde x$).
This formal relation implies a relation between correlators of vertex
operators on a flat background and thus a  $space-time$ symmetry of the
string $S$-matrix.  The off-shell extension of the latter -- the low-energy
effective action
is  then  invariant under the simultaneous  duality transformation
of the `massless' string background fields and interchanging of
the momentum and winding modes.
In particular, the duality maps   isometric string backgrounds which solve the
effective string equations
into other isometric solutions. In the case when the direction  of the isometry
is compact the two dual  solutions correspond to   the same  conformal field
theory,
i.e. are equivalent from the string theory point of view.

{}From a more general point of view, duality  provides  a relation between
processes
at small and large scales and should be  closely connected with the existence
of the
fundamental string scale $L_{Planck}\sim \sqrt{{\alpha}^{\prime}}$, a
 ``stringy'' uncertainty principle
($\Delta x\sim {E}^{-1} +{\alpha}^{\prime}E$) and  existence of critical values
of field strengths and temperature. The duality is likely to  play  an
important role in
understanding  string theory predictions for black hole physics and cosmology
(see, e.g., \refs{\vaf,\giv,\dvv}).
In the simplest case of the torus compactification the duality manifests itself
as a symmetry of the spectrum of states under the interchanging of the momentum
and winding modes and inverting the radius of the torus
$
M_{nm}^{2}\left(r\right)={n^{2}}{r^{-2}}+
{m^{2}r^{2}}{{\alpha^{\prime}}^{-2}}+\ldots ,\ \
\tilde{r} = {\alpha}^{\prime}/r\quad .
$
If the one-dimensional torus of radius $r$ is parametrized by the
periodic coordinate $x\left(\tau , \sigma\right)=r \theta $ then  introducing
the dual string
coordinate,
$\tilde{x}\left(\tau ,\sigma\right)=\tilde r \tilde \tt ,\ \
\partial_{a}x=\epsilon_{ab}
\partial^{b}\tilde{x}  , \  $
 $x=x_{0}+p\ \tau + \tilde{p}\  \sigma +\ldots $,$\  \tilde{x}=
\tilde{x}_{0}+\tilde{p}\ \tau + {p}\sigma +\ldots $, $\ \ \  p=
\sqrt{{\alpha}^{\prime}}n/r$, $\  \tilde{p}=mr/\sqrt{{\alpha}^{\prime}}$,
one can understand  duality as  a  symmetry under which a process
in the $x-$space (torus of radius $r$) corresponds to a dual process in the
$\tilde{x}-$space (torus of radius $\tilde r = \a'/r$) with momentum
and winding modes interchanged. The picture of momentum and winding modes
which correspond to the  same string vacuum but   move in two  different dual
geometries
generalizes to the case of non-trivial curved geometries.


A fundamental feature of string theory is that even
a  large-scale approximation to the geometry is described not only by the
metric but also by the antisymmetric tensor and  the  scalar dilaton field.
The action of a string  in  such a background  is
 \eqn\stri{ I= {1\over { 4 \pi \a' }} \int d^2 z \sqrt {\g} \ [  \del_m x^\m
\del^m x^\n G_{\m
\n}(x)\  + \ i \ep^{mn} \del_m x^\m \del_n x^\n  B_{\m\n}(x) \ } $$  +  \ \a'
R^{(2)}  \p (x)
\ ] \  .$$
$\g$ is a  world sheet metric
which decouples once the background fields satisfy the string  field equations
 which can be derived from the effective action \act.
The classical string probe follows the `geodesic' equations
corresponding to the action of a string  in  such a background  \stri.
One of the consequences  of the  extended nature of a string is that
though a point-like (or zero mode) string state
follows the geodesics of the metric,  other string modes
do feel the antisymmetric tensor background.
First-quantised test string  also feels the dilaton background \refs{\frts}.

An important lesson of the discussion of duality
is that the  \at  and dilaton  play the roles
  very similar   to that of the metric.
In particular, some  metric properties of a background can be  exchanged for
some properties of the \at and/or dilaton  background.
 If  \stri\  is invariant under an abelian  isometry, $x^1\ra x^1 + a$
 then
choosing  the coordinates  $ \{ x^\m \} = \{ x^1 \equiv \tt , \  x^i \} $
 one finds  the
dual action $\tilde {I}$  (1)
for  $ \{ {\tilde x}^\m \} = \{ {\tilde x}^1 \equiv \tilde \tt  , \  x^a \} $
  with ($ M \equiv G_{11}$)
\eqn\duali{ {\tilde G}_{11} = M^{-1} \  , \ \ {\tilde G}_{1a} =  M^{-1} B_{1a}
\
, \ \ {\tilde G}_{ab} =  G_{ab} - M^{-1} ( G_{1a} G_{1b}- B_{1a} B_{1b})\    ,
}  $$
{\tilde B}_{1a} =  M^{-1} G_{1a} \
, \ \ {\tilde B}_{ab} =  B_{ab} - M^{-1} ( B_{1a} G_{1b}- G_{1a} B_{1b})\  ,
 \ \   {\tilde \phi } = \phi - \ha  \ln M  \ .$$
These transformation rules generalise to the case of  $N$ abelian isometries.
If $(G,B,\p)$ is a solution of the string equations
then $(\tilde G , \tilde B,\tilde \p)$  is also a solution (in general,
corrected by $\a'$ terms \tsmpl)
 and (assuming $\tt$ is periodic) the two dual string  actions  correspond to
the $same$
CFT which  is behind a given string background  \rocver.

\subsec{Duality between curved and flat space-times: $D=3$ examples }


 To illustrate the above remarks let  us now consider the backgrounds with  a
covariantly  constant null Killing vector \refs{\guv,\host,\tsnul} \
 which provide a possibility to  study duality   exactly to all orders in
$\a'$.    There exists a simple class of  such `plane-wave' backgrounds
($i,j=1,...,N$)
\eqn\fou{ ds^2= G_{\m\n} dx^\m dx^\n = dudv + G_{ij}(u) d\tt^id\tt^j + dx^a
dx_a \ , } $$ \ \ \ B_{\m\n} =  (0, B_{ij} (u)) \ ,  \ \ \ \ \p=\p(u) \  ,   $$
 where  $\tt^i$ are periodic  and $x^a$ are `spectator' flat coordinates (which
we shall ignore in what follows). The fields in \fou\  are subject to  just one
 string field equation or the
conformal invariance constraint
 and represent  {\it exact}
classical solutions of string theory  which transform into each other under the
full duality group  $O(N,N|Z)$ \kltspl.

 In the case of the {\it flat}  space   ($G_{ij}=\d_{ij}$) the conformal
invariance  of (1) is  maintained by  the balance of
contributions
of  the
antisymmetric tensor $and$ dilaton fields.  Some   duality transformations
may rotate  a  flat metric into   curved backgrounds.
For example,   a   curved   ($R_{iuju}\not=0$) four-dimensional
solution
$\ G_{ij}=\pmatrix{1&f\cr f& 1+f^2\cr}\ , \ \ \ B_{ij} =0 \ , \ \ \  f=f(u) \ ,
\ \ \p = \p(u)  ,  \  $
is related by  the    duality  rotation
 in
the  $\tt^1$-direction  to  the flat metric one
$ \tilde G_{ij} = \d_{ij} \ , \ \ \tilde B_{ij} = - f\epsilon_{ij} \  , \ \
\tilde \p= \p  \ . $
Thus a curved space  is equivalent (from the string theory point of view)
to the
 one with a  flat  metric but non-trivial antisymmetric tensor and dilaton,
i.e.  both  backgrounds give different space-time
representations of the {\it same }  exact string  solution.


To clarify further  the meaning of the relation of flat spaces to curved ones
under duality let us now consider  an even simpler example with $B_{ij}=0$:
a flat $D=3$ background
\eqn\fiv{  ds^2= -dt^2 + dx^2 + (x-t)^2 d\tt^2 =  dudv + u^2 d\tt^2 \ , \  \ \
\ \p=\p_0=\const  \ . }
The dual background
\eqn\six{ ds^2= dudv + u^{-2} d{\tilde \tt}^2 \ , \  \ \ \ \tilde \p=\p_0 -
\ln\ u \ ,\ }
 has a non-zero curvature,
$ \   R_{u\tt u \tt }= -2 u^{-4}$ and is also an exact solution.
Though flat, the background \fiv\ (called `null orbifold' in \hht) is quite
non-trivial (it does not have a conical singularity at $u=0$).
As in the case of the  flat torus it  is possible to demonstrate explicitly
 that \fiv\ and \six\  represent the $same$  string solution,  i.e. the $same$
conformal theory. Representing  the metric \fiv\  in flat coordinates
$ds^2 = - 2 dUdV + dX^2$ with the identification
$(U,V,X)=(U,V+wX+\ha w^2 U,X+wU)$, $ w= m \pi, m=0,1,2,...,$
it is easy to solve the corresponding string equations
and to show that in addition to the standard momentum  modes there are winding
modes \klits.  Quantising the string canonically
and replacing  the winding number $m$ by
$-{i\over 2}{\del \over\del \tilde\theta}$  one finds  for the zero mode parts
of the  standard  constraints on the string states
\eqn\sev{ (L_0^+-L_0^-)T={1\over
4}{\del ^2\over\del \theta\del \tilde\theta}T=0 \ ,  \ \  T(u,v, \tt, \ttt) =
T(u,v, \tt) + \tilde T (u,v, \ttt) \ , }
\eqn\eig{  {2\over \a'}(L_0^++L_0^--2)T =\big( 2{\del ^2\over\del  u\del
v}-{1\over u^2}{\del ^2
\over \del \theta^2}-{u^2\over 16 \a'^2}{\del ^2\over \del \tilde\theta^2}
+{1\over u}{\del \over \del  v}-{4\over\a'} \big) T =0  \ .  }
Eq.(11) restricted to a solution of the first constraint  takes the form
of a  Klein-Gordon-type equation in a metric-dilaton background
$$ \   (- D^2  + 2 D^\m \p \del_\m  - {4/\a'}) T =0 \  .    $$
We conclude  \klits\  that  while the momentum  point-like state $T(u,v,\tt)$
propagates in the original flat  background \fiv\   the winding
state  $\tilde T (u,v, \ttt)$  `feels'  the dual  curved background \six.
The backgrounds \fiv\ and \six\ thus represent two  complementary   images of
the same  string geometry as  `seen'  by different point-like string states.

It is  useful compare  \fiv\   with the standard   flat $D=3$ background
\eqn\sws{ ds^2 = -dt^2 + dx^2 + dy^2 =  -dt^2 + dr^2 + r^2 d\tt^2\  ,  \  \ \
\p=\p_0 , \ }
which is related by the leading-order duality  to  (see also
\refs{\Ve,\rocver})
 \eqn\ppp{   ds^2= - dt^2 + dr^2  + r^{-2} d\ttt^2 \  , \  \ \  \  \tilde
\p=\p_0 - \ln r   \  .  }
The dual metric is curved and singular; it   solves the leading-order conformal
invariance equations
but is corrected by $\a'$-terms  so that in contrast to the above plane wave
example  (where  both the original and the duality transformed backgrounds
are known explicitly)
here one is unable to make definitive
conclusions, e.g.,  about the  fate of the $r=0$ singularity.
Moreover, the string spectrum on this flat background  does not contain genuine
winding modes so that one cannot   reproduce the dual background
as  a geometry felt by some string modes on  the original background (as was
the case for the torus and the null orbifold \fiv).

Another instructive  example of a flat  $D=3$ space with a  curved dual
counterpart is provided by  a plane-wave type one (but now   outside the  class
 \fou)
\eqn\nin{ ds^2= d\tt dt  + x d\tt^2 + dx^2  \ , \ \ \   \ \p=\p_0  \ ,   }
where $\tt$  is  assumed to be periodic.
This is the simplest example of a `$K$-model'.
Performing the duality transformation with respect to $\tt$ we get a
background  with curved metric and non-constant  antisymmetric tensor and
dilaton  described
 by the following sigma model (which is the simplest example of a $F$-model)
\eqn\ten{ \tilde
I= {1\ov \pi \a'} \int d^2 z  ({x\inv }  \del u \bd v +  \del x \bd x  )
\ , \ \    \ \tilde \p = \p_0 -  \ha \ln x \ , \ \ \ \  u= \ttt - t, \ \ v  =
\ttt + t \ .  }
The space \nin\ is flat and regular (`null manifold')  but has rather bizarre
causal properties.  It  appears that the acausality
of the metric \nin\ is lost in the process of the duality transformation since
the metric in \ten\ $$ds^2=  {x}\inv  dudv +  dx^2  =   {x}\inv(d\ttt^2 -
dt^2)+
dx^2 $$
  does not have  closed time-like geodesics.
However,  a   peculiar causal  structure of  \nin\ is now   reflected  in the
presence of the time-like component of the antisymmetric tensor
in \ten. The causal structure of the string geometry is thus  represented in
different  ways in its  dual images.

As in the case of the flat torus and the null orbifold \fiv\ it is possible to
identify the dual background \ten\ as the geometry  `probed' by the
winding modes
of the string spectrum  corresponding to the  flat space \nin\ \klits. Solving
the string equations using flat coordinates for \nin\ and quantising the theory
 one finds the expressions for the constraints which are similar to \sev,\eig\
and concludes that  while the momentum  zero-mode string state $T(\tt,t,z)$
satisfies the Klein-Gordon equation \ten\ corresponding to the  original flat
background \nin\ the winding  state $T(\ttt, t, z)$
propagates in the curved metric-dilaton background \ten.

\newsec{Exact solutions corresponding to gauged WZW models}

The general bosonic \sm describing string propagation in a `massless'
background is given by \stri, i.e. in conformal gauge
\eqn\mod{  I= {1\ov \pi \a' } \int d^2 z L\ , \ \ \ L=  (G_{\m\n} +
B_{\m\n})(x)\ \del x^\m \bd x^\n
+ \a'{\cal R}\p (x)\ ,  }
where $G_{\m\n}$ is the metric,  $B_{\m\n}$ is the antisymmetric tensor and
$\p$ is the dilaton
(${\cal R}$ is related to the  worldsheet metric $\g$  and its
scalar curvature by  $ {\cal R} \equiv \fourth \sqrt \g R^{(2)}$).
The action of the ungauged WZW model for a group $G$ \wit\
 \eqn\unga{  I=kI_0\ , \ \ \   I_0 \equiv  {1\over 2\pi }
\int d^2 z  \Tr (\del g^{-1}
\bd g )  +  {i\over  12 \pi   } \int d^3 z \Tr ( g^{-1} dg)^3   \ ,
}
can be put into the general \sm form \mod\
by introducing the coordinates on the group manifold. Then
$G_{\m\n}$ is the group space metric,  $H_{\m\n\l} = 3 \del_{[\m}B_{\n\l]}$
is the parallelising torsion  and dilaton $\p=\p_0=\const$ (and $k=1/\a'$). In
a special scheme (where the $\beta$-function of the general model \mod\ is
proportional to the generalised curvature)
the  resulting \sm is finite  at each order of $\a'$-perturbation theory
\refs{\bra,\mukh} and corresponds to the well-known current algebra CFT \wit.

The action of the gauged WZW model  \gwzw\
\eqn\gau{
I (g,A) = kI_0(g)  +{k\over \pi }
 \int d^2 z \Tr \bigl( - A\,\bd g g\inv +
 \bar A \,g\inv\del g   + g\inv A g \bar A  - A \A \bigr)
   \ ,  }
is invariant under  the    vector  gauge transformations
with parameters taking values in a subgroup $H$ of $G$. Parametrising  $A$ and
$\A$ in terms of $\  h$ and $\bh$  from $H$,
$ A = h \del h\inv \  , \  \ \A = \bh
\bd \bh\inv $ one can  represent \gau\   as the
difference of the two manifestly gauge-invariant terms: the ungauged  WZW
actions
corresponding to the group
$G$
and  the subgroup $H$,
\eqn\gact{ I (g,A) =k I_0 (h\inv g \bh ) -  kI_0 (h\inv \bh)  \  . }
This representation  implies that the  gauged WZW  model corresponds to  a
conformal
theory (coset CFT \refs{\GKO,\bha}).
Fixing a  gauge on $g$ and changing the variables to $g' = h\inv g{\tilde h} ,
\
 h' = h\inv {\tilde h}$  we get a \sm on the group space $G\times H$ which is
conformal to all orders in a particular  `leading-order'  scheme.
That means that the 1-loop  group space solution remains  exact solution in
that scheme.
Replacing \gact\ with  the `quantum' action with renormalised levels
$k\ra  k + \ha c_G$ and $k\ra  k + \ha c_H$  does not change this conclusion.
This replacement   corresponds to  starting with the theory formulated in
the `CFT' scheme  in which, e.g.,  the  exact central  charge of the WZW model
is reproduced by the first non-trivial correction \refs{\tspl, \tssfet}
 and the metric $( k + \ha c_G ) G_{\m\n}$ is the one that appears in the CFT
Hamiltonian $L_0$  considered as a Klein-Gordon operator.

\subsec{`CFT' and `leading-order' schemes}
To obtain the corresponding  \sm in the `reduced'  $G/H$
configuration space (with coordinates being parameters of gauge-fixed $g$) one
needs to integrate out $A,\A$ (or, more precisely, the WZW fields $h$ and
${\tilde h}$).  This is a non-trivial step and the form of the result depends
on a choice of a scheme in which the original
 `extended' $(g,h,{\tilde h})$  WZW theory is formulated.

Suppose first   the latter is taken in the leading-order  scheme
with the action \gact.
Then the   result of integrating out $A,\A$
and fixing a gauge takes the form of the \sm \mod\ where the \sm metric and
dilaton are then given by (see \horts\ and refs. there)
\eqn\resm{ G'_{\m\n} = G_{\m\n}  - 2 \a' \del_\m \p \del_\n \p \ , \ \ \
\p=\p_0  - \ha \ln \det F \ . }
$G_{\m\n}$ is the metric obtained by solving for $A,\A$ at the classical
level
and $\p_0$ is the original constant dilaton.
Since the  $\a'$-term in the metric can be  eliminated by a field redefinition
we conclude that there exists a {\it leading-order scheme}  in which the
leading-order  gauged WZW
\sm
background $(G,B,\p)$ remains an  exact solution. The leading-order scheme
for the ungauged  WZW \sm   is thus   related to the leading-order scheme for
the gauged  WZW $\s$-model
 by an extra $2 \a' \del_\m \p \del_\n \p$ redefinition of the metric.
This   provides a general explanation for the observations in \refs{ \tspl,
\tssfet} about the existence of a leading-order scheme
 for  particular $D=2,3$ gauged WZW models.

If instead we start with the  $(g,h,{\tilde h})$  WZW theory in the CFT scheme,
i.e
 with the action
\eqn\gactt{I(g,A) = (k+ \ha c_G)\big[ I_{0} (h\inv g{\tilde h})- {k+ \ha
c_H\ov k+ \ha c_G}
I_{0} (h\inv {\tilde h})\big] \ , }
then the resulting  \sm couplings will explicitly depend on $1/k\sim \a' $
(and will agree with the  coset CFT operator approach results
\refs{\dvv,\bsfet,\tsnp}).
While in  the WZW model the transformation from the  CFT to the leading order
scheme is just a simple rescaling of couplings,  this  transformation  becomes
non-trivial    at the level of gauged WZW $\s$-model.   It is the  `reduction'
of the configuration space
resulting  from   integration over the gauge fields
$A,\A$  that is
 responsible for  a  complicated form of the transformation law between the
`CFT' and `leading-order' schemes in the gauged WZW $\s$-models  (in
particular,  this transformation  involves  dilaton terms of  all orders in
$1/k$, see  below).\foot{An exception is provided by some  $\s$-models
obtained by nilpotent
gauging:
here the second term in \gactt\ is absent by construction \klts. The background
fields do not receive non-trivial $1/k$ corrections even in the CFT scheme,
i.e.  the relation
between the leading-order and CFT schemes is equivalent to the one  for the
ungauged WZW model.
The  same is true for the $D=3$ \FM or the extremal limit of the $SL(2,R)\times
R/R$ coset.}

The basic example is  the $SL(2,R)/R$ gauged WZW model (or $D=2$ black hole)
\refs{\witt,\brsne,\cres}.
 The  (euclidean) background in the leading-order scheme
\eqn\exann{ ds^2 =    dx^2 + \ { {\tanh}^2 bx }   \ d{\t}^2   \ ,  \ \ \
 \p = \p_0  -  {\ln \ch }  bx  \ ,  }
is related to the  background in the CFT scheme \dvv
\eqn\exann{ ds^2 = dx^2 + \ { {\tanh}^2 bx \ov 1 \ - {\ p \ }
{\tanh}^2 bx }   \ d{\t}^2   \ ,  \ \ \  }  \eqn\eeee{
 \p = \p_0  -  {\ln \ch }  bx   -
\fourth  \ln \bl( 1 \ - \ p \ {\tanh}^2 bx )\ ,  }
\eqn\exxxx{  \  p\equiv {2\ov k} \ , \ \
\a'b^2 = {1 \ov k-2}\ , \ \ \ D - 26  + 6\a'b^2 =  {3k\ov k-2} -1 -26  =0  \ ,
\ \ D=2\  , }
by the following local covariant redefinition \tspl
\eqn\cftf{   G^{(lead)}_{\m\n}  = G^{(cft)}_{\m\n}  - { 2\a' \del_\m \p \del_\n
\p  \ov 1  + \ha \a' R  }      + {2\a' (\del \p)^2  G_{\m\n}\ov 1 + \ha \a' R}
\ , }
$$  \p^{(lead)}=\p^{(cft)}  - \fourth  \ln ( 1  +
{\textstyle {1\ov 2}} \a' R  ) \ .   $$
It should be emphasized that the two backgrounds related by \cftf\
describe the same  string geometry since the probe of the geometry is the
tachyon field
and the tachyon equation  remains the same  differential equation
implied by coset CFT (even though it looks different being expressed in terms
of
different $G$ and $\p$).

Similar remarks apply to the
 a  particular  $D=3$ solution  -- the `charged black string'
corresponding to $[SL(2,R)\times R]/R$ gauged WZW model \hoho.
Its form in the leading-order scheme can be obtained by
making a duality rotation of the neutral black string, i.e. the direct product
of the $D=2$ black hole and a free scalar theory.

\subsec{$D=4$ solutions}

In general, the backgrounds obtained from  the gauged WZW models
have very few (at most,  abelian) symmetries and thus cannot directly describe
 non-trivial $SO(3)$-symmetric  backgrounds which are of interest in connection
with
 $D=4$ cosmology and black holes.
Let us mention the explicitly known $D=4$  solutions with $(-,+,+,+)$
signature.
A class of axisymmetric black-hole-type and anisotropic cosmological
backgrounds is found
by starting with $[G_1\times G_2]/[H_1\times H_2]$ gauged WZW model with
various combinations of $G_i=SL(2,R)$ or $ SU(2)$ and $H_i= R$ or $U(1)$
\refs{\horv,\kou,\napwit,\givpas}.
The leading-order form of these backgrounds can be obtained by applying the
$O(2,2)$ duality transformation
to the direct product of the Euclidean and  Minkowski $D=2$ black holes or
their
analytic continuations.
The stationary black-hole-type background is \horv
\eqn\dualih{ds^2=  -{g_1 (y) \ov g_1 (y)g_2 (z)  - q^2 }\
dt^2 +
 {g_2 (z)\ov g_1 (y)g_2(z)  - q^2} \ dx^2   + dy^2 + dz^2 \ , }
$$B_{tx }= {q \ov  g_1 (y) g_2 (z) - q^2}   \ , \ \ \
\phi= \p_0 -\ha \ln \bl(\sinh^2y \sinh^2z  [ g_1(y) g_2(z) - q^2]\br)  \ ,  $$
where $g_1= \coth^2 y, \ g_2= \coth^2 z$.
The cosmological background  \napwit\ can be obtained by the analytic
continuation
and renaming the coordinates
\eqn\dualiw{ds^2= -dt^2 + dx^2 +  {g_1 (t) \ov g_1 (t)g_2 (x)  + q^2 }\
dy^2 +
 {g_2 (x)\ov g_1 (t)g_2(x)  + q^2} \ dz^2    \ , }
$$B_{yz }= {q \ov  g_1 (x) g_2 (t) + q^2}   \ , \ \ \
\phi= \p_0 -\ha \ln \bl(\cos^2t \cosh^2 x  [ g_1(t) g_2(x) + q^2]\br)  \ ,  $$
where $g_1= \tan^2 t, \ g_2= \tanh^2 x$.
A  non-trivial (`non-direct-product') $D=4$ background without isometries is
obtained  from  $SO(2,3)/SO(1,3)$ gauged WZW model \barsf.

Two other classes of $D=4$ Minkowski signature   `coset' solutions will be
discussed in more detail
in the following sections 5.1 and 6.2.  The first includes  plane-wave-type
backgrounds
which are obtained from non-semisimple versions of  $R\times SU(2)$ and the
above
$[SL(2,R)\times SL(2,R)]/[R\times R]$, etc. `product' cosets \refs{\napwietc,
\sfeets}.
The second contains $F$-models (backgrounds with 2 null Killing vectors)
 which follow from nilpotent gauging of rank 2 maximally non-compact groups
\klts.


\newsec{Solutions with covariantly constant null Killing vector}

The string action in a background with a covariantly constant null Killing
vector
can be represented in the following general  form ($i,j=1,...,N$)
\eqn\str{   L =   \del u \bd v +
  K(u,x)\del u \bd u  +  A_i(u,x) \del u  \bd  x^i  + \A_i(u,x) \bd u \del  x^i
}
$$+  (G_{ij} + B_{ij})(u,x)\ \del x^i \bd x^j  \  +  \ \a'  {\cal R}  \p (u,x)\
 , $$
$$I= {1\over {  \pi \a' }} \int d^2 z \  L  \ , \ \ \
\ \ \ \   {\cal R} \equiv \fourth \sqrt \g R^{(2)} \ .  $$
$K, A_i, \A_i$ can be eliminated (locally) by a coordinate and  $B_{\m\n}$-
gauge transformation \refs{\brink,\tsnul}
so that the general form of the Lagrangian is
\eqn\sstr{ L =   \del u \bd v
 +   (G_{ij} + B_{ij})(u,x)\ \del x^i \bd x^j  \  +  \ \a'  {\cal R}  \p (u,x)
\  ,  }
with $K, A_i, \A_i$ been now `hidden' in a  possible general coordinate and
gauge transformation. If the `transverse' space is trivial (for fixed $u$)
one may use \sstr\  with `flat' $G_{ij}, B_{ij}$ taken in the most general
frame, or may
choose a special frame where $G_{ij}=\delta_{ij}, B_{ij}=0$ and  then   go back
to  \str\
(which may be preferable for a global coordinate choice).

There are two possibilities to   satisfy the conditions of conformal
invariance.
 The first   is realised in the case when  the `transverse' theory
is conformally  invariant. The second option is to include the term linear in
$v$ to the dilaton field \tsnul.
Then the conformal invariance  conditions are satisfied provided
the `transverse' couplings $G,B$ depend on $u$ according to the standard RG
equations
 with the $\b$-functions of the `transverse' theory.  Since these
$\b$-functions are not known in a closed form (except for some supersymmetric
\sm cases \tsnuu)
one is unable to determine the explicit all-order form of the solution.
For this reason  here  we shall consider only the first possibility.

\subsec{Plane waves}

The simplest  special case is that of the `plane-wave' backgrounds for which
$G,B,\p$ in \sstr\ do not depend on $x$
(i.e., in particular, the transverse metric is flat).
The conformal invariance condition then reduces to one equation
\eqn\eeeq{ \   -{1\over 2} G^{ij}  {\ddot G}_{ij} + \fourth  G^{ij} G^{mn}(\dot
G_{im}\dot G_{jn}
 -{\dot B}_{im}{\dot B}_{jn} )
 + 2 \ddot \p =0\  \  .  }
 $G_{ij } (u), B_{ij} (u) $  and $\p (u)$ satisfying \eeeq\
  represent  {\it exact}
classical solutions of string theory  which transform into each other under the
full duality symmetry  group  $O(N,N|Z)$ \kltspl.

A subclass of  such  models  admits a coset CFT interpretation in terms
of gauged WZW models for  non-semisimple groups \refs{\napwietc, \sfeets}.
The latter can be obtained from standard  semisimple gauged WZW models by
taking special singular limits. The singular procedure  involves a coordinate
transformation and a rescaling of $\a'$ and is carried out directly at the
level of the string action \sfeets.\foot{It is similar  to  the  transformation
 in \penr\ which maps
any  space-time  into some plane wave.}
As a result, one is able  not only to obtain the plane-wave background fields
but also to relate  the corresponding CFT's (i.e. to construct a
`non-semisimple' coset CFT  from a `semisimple' coset CFT)
 and thus to give
a CFT description to this  subclass of plane-wave solutions.

In particular,  a  set of  $D=4$ plane wave  solutions that can be obtained in
this way from
gauged  $[SU(2)\times SL(2, R)]/[U(1)\times R]$ WZW models is given by \sfeets\
(cf. \dualiw)
\eqn\duali{ds^2= dudv +  {g_1 (u') \ov g_1 (u')g_2 (u)  + q^2 }\
dx_1^2 +
 {g_2 (u)\ov g_1 (u')g_2(u)  + q^2} \ dx_2^2  \ , }
$$B_{12 }= {q \ov  g_1 (u') g_2 (u) + q^2}   \ , \ \ \
\phi=\p_0  -\ha \ln \bl(f_1^2 (u')f_2^2 (u) [ g_1(u') g_2(u) + q^2]\br) \ ,  $$
where $u'= au + d \ $ ($a,d=\const$) and the functions $g_i, f_i$ can take any
pairs of the following values
 \eqn\hhh{\eqalign{&g(u) = 1\ , \ \ u^2\  ,  \ \  \tanh^2 u \ , \ \  \tan^2 u \
, \ \
 u^{-2}\ , \ \ \coth^2 u\ , \ \ \cot^2 u \ , \cr
 &f(u) = 1\ , \ \ 1\ , \ \ \ \  \cosh u\ , \ \ \ \ \cos\ u \ , \ \
\ u\ , \ \ \ \sinh u \ , \ \  \  \sin u \ . \cr } }
A particular case is  $g_1=1, \ g_2 =u^2$
(this    background   is dual to  flat space). Another special case is the
 $E^c_2$  WZW model of Nappi and Witten \napwietc\foot
{ $E^c_2$ is  a central extension of the Euclidean  algebra  in two dimensions
admitting  a non-degenerate invariant bilinear form.}
\eqn\acd{L = \del v\bd u + \del x_1 \bd x_1
+\del x_2 \bd x_2 + 2\ \cos\ u\ \del x_1 \bd x_2 \ , }
which   can be obtained by a singular limit from
 the WZW action for
$SU(2)\times R$.

For $B_{ij}=0$  \sstr\ can be transformed into the familiar form (i.e.  \str\
with $G_{ij}=\delta_{ij}, \ A_i=\A_i=B_{ij}=0$)
\eqn\sstry{ L =   \del u \bd v +  K(u,x) \del u \bd u
 +   \del x^i \bd x_i  \  +  \ \a'  {\cal R}  \p (u)
\  .   }
The fact that this model  has a covariantly constant null vector
can be used to give a simple geometrical argument that
   leading order solutions  are exact
when  $G_{ij} =\d_{ij}, \ B_{ij} =0$, and $\p$ depends only on $u$ \host.
This is because the curvature contains two powers of the
constant null vector $l$, and derivatives of $\p$ are also
proportional to $l$. One can thus show that all
higher order terms in the equations of motion vanish identically.
The one-loop conformal invariance condition
\eqn\tacc{
 - \ha  \del^i\del_i K    +
2 \del_u^2  \p =0 \  , }
is then the exact one and is solved by  the standard plane-wave ansatz
\refs{\guv,\host}   $K= w_{ij} (u) x^i x^j $.
A different rotationally symmetric solution  exists for $\p=\const$
\eqn\ffss{ K = 1 + {M\ov r^{D-4}}  \ , \ \ D> 4 ; \ \ \
K= 1- M\ln\ r \ , \ \ D=4; \ \ \
 \ r^2 \equiv  x_ix^i > 0  \  , \ \ \
\p=\const\ . }
This background is dual to the  fundamental string  solution \hhs\ and
describes a
string boosted to the speed of light.

\subsec{Generalised plane waves}
If one sets $G_{ij}=\d_{ij}, \ B_{ij}=0$ in \str\  but keeps $A_i, \A_i$
non-vanishing the conditions of conformal invariance
take the form \tsnul\
\eqn\con{\del^j F_{ij} =0 \ , \ \ \ \  \del^j \bar F_{ij} =0 \ ,  }
\eqn\coni{- \ha  \del^2 K  +   F^{ij} \bar F_{ij}
+   \del^i{\del_u( A_i + \A_i)}
+ 2 {\del^2_u\p} }
$$ +   O\big({\a'}{}^s (\del^s F)^2 , \ {\a'}{}^s(\del^s F)^2 ,\
{\a'}{}^{s+k}\del^s F \del^{k}\bar F\big) =0 \ . $$
It is clear that one-loop expression simplifies if $A_i=0$ or $\A_i=0$.
Similar simplification happens at the two-loop level \duval.
Plane waves with $\A_i=0$ were shown to preserve (half of) supersymmetry
and to have certain $\a'$-corrections to vanish in the context of superstring
effective action \bergsh.
In fact, it is possible  to show  \hrts\ that such `chiral'  plane wave
backgrounds
with $\A_i=0$ (or $A_i=0$)
\eqn\stre{   L =   \del u \bd v +
  K(u,x)\del u \bd u  +  A_i(u,x) \del u  \bd  x^i + \del x^i \bd x^i    +
\a'  {\cal R}  \p (u,x)\  , }
 are exact solutions of bosonic string  theory  provided
$K$,$A_i, \p$ satisfy the one-loop conformal invariance conditions,
\eqn\sses{- \ha  \del^2 K
+   \del^i\del_u A_i
+ 2 {\del^2_u\p} =0 \ , \ \ \ \del^j F_{ij} =0 \ , }
i.e. all higher order terms in \coni\ are actually of the $F\bar F$-structure
  $O\big({\a'}{}^{s+k}\del^s F \del^{k}\bar F\big)$, i.e. vanish for $\A_i=0$.

\subsec{\KM }

Another special case of \str\ is found when $K,G_{ij}, B_{ij},\p$ do not depend
on $u$.
This is  a generalisation of the plane wave solutions called the `$K$-model' in
\horts
\eqn\mofk{ L_{K}= \del u \bd v +  K(x) \ \del u \bd u  + (G_{ij} +
B_{ij})(x) \ \del x^i \bd x^j     + \a'{\cal R}\p (x)\ .  }
In general, suppose that the transverse space is  known to be an exact  string
solution  in some scheme. Then the model \mofk\ is  be conformal for $K=0$. But
the curvature of the metric with $K \ne 0 $ is equal to the curvature
of the metric with $K=0$ plus a term of the form $(\na \na K)\ l\ l $.
Unfortunately, this can
result in nontrivial corrections to the equations of motion at
each order of $\a'$. These corrections will be linear in $K$, so one
learns that  the
exact equation for $K$ will also be linear.
The exact form of the equation for $K$ turns out to be the following \horts
\eqn\taac{  \
- \o K    +  \del^i \p \del_i K+ 2 \del^2_u\p
=0 \ , \ \ \ \  \o = \ha  \na^2   +  O(\a') \ ,  }
where $\o$ is the scalar anomalous dimension operator
which in general contains $(G_{ij},B_{ij})$-dependent corrections to all
orders in $\a'$ (only  a few leading $\a'^n$-terms in it are known
explicitly, see  e.g. \tspl\ and refs. there).

Given an exact  string solution $(G_{ij},B_{ij}, \p )$,   in general,  we would
still be unable to determine the exact expression for $K$ because of the
unknown higher order terms in \taac.
There are, however, special cases when this is possible.
The simplest one is that of the flat transverse space with the  dilaton
 being linear in  the coordinate $x$,  $\ \p = \p_0(u)  +
b_i(u)x^i \ $ which is an obvious generalisation of the plane wave case \tacc.
It is possible to  obtain more interesting new exact solutions when the
CFT behind the `transverse' space solution $(G_{ij},B_{ij} ,\p)$ is
nontrivial but still known explicitly \horts.
In fact, in that case the structure of the  `tachyonic' operator
$\o$ is determined by the  zero mode part of the
CFT Hamiltonian, or $L_0$-operator.
Fixing a particular scheme (e.g.,  the `CFT' one
where $L_0 $ has the standard Klein-Gordon form with the dilaton term)
 one is able, in principle,  to  establish  the form
of  the background fields $(G_{ij},B_{ij} ,\p)$
$and$ $K$. This produces  `hybrids' of  gauged WZW and plane wave solutions.

To obtain a {\it four dimensional} hybrid
 solution \horts\ one must start with a two
dimensional conformal $\s$-model. Essentially the only non-trivial  possibility
is
the   $SL(2,R)/U(1)$
gauged WZW model which describes the two dimensional euclidean black hole,
i.e. the resulting solution  will
be
the $generic$ $D=4$ $K$-model.
To construct this new solution we must
use all the zero-mode information provided by the
$SL(2,R)/U(1)$  coset: the exact metric and  dilaton   and the  form of the
tachyon equation. This will give us  the all-order form of all the
functions in the $D=4$  model.

The exact background fields
of the  $SL(2,R)/U(1)$ model in the CFT  scheme were  given in \exann,\eeee.
In the CFT scheme the tachyonic equation has the
  standard uncorrected form,  so that
 the  function $K(x)$ must satisfy
\eqn\tach{ - \ha  \na^2  K  +   \del^i \p  \del_i K=
- {1\ov 2\sqrt G \e{-2\p} } \del_i (\sqrt G \e{-2\p} G^{ij} \del_j) K =  0 \ .
\  }
A particular solution of \tach\  with $K$ depending  only on $x$ and not on
$\t$
 is
\eqn\taa{ K= a  + \  m  \ \ln \tanh bx  \  . }
The constants $a, \ m $ can
 be absorbed into  a redefinition of $u$ and $v$,
 so that the full exact $D=4$ metric is \horts
\eqn\mix{ds^2 = dudv + \  \ln \tanh bx \ du^2  + dx^2 +    { {\tanh}^2 bx \ov 1
\ - {\ p \ } {\tanh}^2 bx }\ d{\theta}^2 \ ,  }
while the dilaton is unchanged.
This metric is asymptotically flat,  being a product of $D=2$ Minkowski
space with
a cylinder
at infinity.

The solution for $K$ \taa\
 is the $same$ in the `leading-order' scheme
where the metric and dilaton do not receive $\a'$ corrections.
The point is that the tachyon operator remains the same  differential operator,
it is only its expression in terms of the new $G, \p$ that changes.
Thus, in the `leading-order' scheme we get the following exact $D=4$ solution
\eqn\mixl{ds^2 = dudv +  \ln \tanh bx \  du^2  + dx^2 +
   { {\tanh}^2 bx }\ d{\theta}^2 \ , \ \ \ \p = \p_0  -  {\ln \ch }  bx \ . }
In addition to the    covariantly constant null vector $\del/ \del v$,
this solution  has
two
isometries  corresponding to shifts of $u$ and $\t$. Hence we can consider
two different types of duals. Dualizing with respect to $\t$  gives
\eqn\mixll{ds^2 = dudv +  \ln \tanh bx \  du^2  + dx^2 +
   { {\coth}^2 bx }\ d{\theta}^2\ ,
 \ \ \   \p = \p_0  -  {\ln\  \sinh \  }  bx \  . }
Starting  with the general
solution for $K$ \taa\ one finds the  $u$-dual background \horts\
which will be discussed in section 6.3.

\newsec{ $F$-model solutions }

The simplest \FM  \refs{\klts,\horts}  describes
a family of backgrounds   with
metric and antisymmetric tensor
characterized by a single function $F(x)$ and dilaton $\p (x)$:
\eqn\ffff{ ds^2 = F(x) du dv + dx_i dx^i\ ,   \qquad B_{uv} = \ha F(x) \ . }
The two functions $F$ and $\p$ depend only
on the transverse coordinates
$x^i$. The leading order string  equations
then reduce to  \klts
\eqn\fmod{ \del^2 F\inv = 2b^i \del_i F\inv\ , \ \ \
  \p = \p_0 + b_i x^i  + \ha  \ln F (x) \ , \ }
where $b_i$ is a  constant vector.
Some of the solutions to \fmod\  were  shown   to
correspond to gauged WZW models where the subgroup being gauged
 is nilpotent \klts (see section 6.2 below). It was
argued that like ungauged WZW models they should not receive non-trivial
higher order corrections even in
the CFT scheme. Though it is unlikely that
 all solutions to \fmod\ can be obtained from a gauged WZW model
one can   show  \horts\ that there
exists  a scheme in which all of these solutions are exact
and receive no $\a'$ corrections. Since the equation for $F\inv$ is linear,
linear combinations of these solutions yield new exact solutions.

One of the most interesting solutions in this class  is the
`fundamental string' \gibb\
 which
has $b_i=0$ and
\eqn\fss{ F\inv ={ 1 + {M \ov r^{D-4}} }   \ , \ \ D>4 \ ; \ \ \
 F\inv ={ 1 - {M \  \ln \ r } } \ , \ \ \  D=4\ , \ \  \ r^2 = x_i x^i\ ,  }
where $D$ is the number of spacetime dimensions.
This solution
describes the field outside of a straight fundamental string located at
$r=0$.

\subsec{General  \FM }
The general \FM  has a curved transverse space, i.e. is defined  as  \horts
\eqn\mof{  L_F=F(x) \del u \bd v +  (G_{ij} + B_{ij})(x)\ \del x^i \bd x^j
+ \a'{\cal R}\p (x)\ .  }
This model is related by leading order duality to \KM \mofk: the dual
of  it  with respect
to $u$ is \mof\ with $F=K\inv$.
The \FM has a large symmetry group. It is invariant under
the  three Poincare transformations
in the $u,v$ plane.  Moreover, it is invariant
 under
the infinite-dimensional symmetry $u\ra u + f(\tau+\s),
\   v\ra v + h(\tau-\s)$, i.e.  it has
two chiral currents.

The all-order conformal invariance conditions for \FM
are satisfied \horts\ provided one is given a
conformal `transverse' theory  $(G', B', \p')$  and
\eqn\summm{  G_{ij} =G'_{ij}  +  \ha \a' \  \del_{i } {\ln } F \  \del_{j}
{\ln} F \ , \ \ \  \p= \p'   + \ha \ln F \ , \
\ \ B_{ij}= B'_{ij} \ ,   }
with  $F$ satisfying
\eqn\taccc{ -   \o' F\inv +  \del^i \p' \del_i F\inv
=- \ha  \na'^2 F\inv    +  O(\a')  +  \del^i \p' \del_i F\inv =0
\ . }
Here $\o'$  is the anomalous dimension operator depending on $G'$.
When   $(G',B,\p')$ correspond to  a  known CFT
this equation
can be written down explicitly to all orders in $\a'$.
Since the relation between $G$ and $G'$ is local  and
the transverse theory  is,  in general,  defined modulo local coupling
redefinitions,  one can argue \horts\ that there
exists a (`leading-order') scheme in which  the exact solution is represented
by
the conformal transverse theory $(G,B,\p')$  and  $F$ satisfying \taccc.

The   simplest  example of the  conformal  transverse model is the flat space
with  a linear dilaton, i.e. the  corresponding \FM  is  represented
(in the leading-order   scheme) by
\eqn\exaav{  G_{ij} =\d_{ij}  \  , \ \ \
\ \  \p= \p_0  + b_i x^i   + \ha\  \ln \  F \ . }
In this  scheme, the exact  form of the equation for the function $F$ is
simply
\eqn\fff{ - \ha  \del^2 F\inv  +  b^i \del_i F\inv =0 \ , }
i.e. the  model
\eqn\fef{  L_{F}=F(x) \del u \bd v +  \del x^i \bd x_i    + \a'{\cal R}
 (\p_0 +b_ix^i +    \ha \ln F ) \  , }
with $F$ satisfying \fff\ is  conformally  invariant to all orders, i.e.
gives an exact string solution.

In this scheme the leading-order duality is exact since the
leading-order  dual to \fef\ is the $K$-model
\eqn\feff{  L_{K}=\del u \bd v +   F\inv (x) \del x^i \bd x_i    + \a'{\cal R}
 (\p_0 +b_ix^i) \  ,   }
which represents an exact  solution if $F$ solves \fff\
(cf.\tacc\ with $\p =0$).
In particular, one concludes that there exists a scheme in which the FS
solution
\fss\
is a classical string solution to all orders in $\a'$.

\subsec{$F$-models  obtained  by nilpotent  gauging of   WZW  models}

The conclusion   about the existence of  a scheme where the
$F$-model \fef\ represents an exact string solution  is consistent with the
result of \klts\  that
the particular $F$-model with
\eqn\klt{  F\inv =  {  \sum_{i=1}^N \ep_i {\rm e}^{  \a_i\cdot x  } }\ , \  \ \
\ \
\p =   \p_0 +   \r \cdot x   +
\ha\  \ln \ F
\   ,   }
can be obtained  from  a $G/H$ gauged WZW
model.
Here  the constants $\ep_i$ take values  $ 0$ or $ \pm 1$, $\ \a_i$
are simple roots of the   algebra of a maximally non-compact
Lie group $G$ of rank $N=D-2 $
 and $\r= \ha
\sum_{s=1}^m  \a_s$ is
 half  of the sum of all positive roots.
 $H$ is  a  nilpotent subgroup of $G$
generated by $N-1$ simple roots (this condition on $H$
 is needed to get  models with one time-like  direction).
The flat transverse coordinates $x^i$ are  the Cartan subalgebra directions.

 The corresponding `null'  gauging \klts\ is based on the  Gauss decomposition
 and  directly applies only to the
groups  with the algebras that are  the `maximally non-compact' real forms of
the classical Lie algebras
(real linear spans of the Cartan-Weyl   basis), i.e.   $sl(N+1,R), \ so(N,N+1),
\ sp(2N,R), \ so(N,N)$. These WZW models can be considered as
natural generalisations of the $SL(2,R)$ WZW model.  For such  groups there
exists a real group-valued Gauss
decomposition
$$  g=  \exp ( \sum_{\Phi_+ } u^\a E_{\a }) \  \exp ( \sum_{i=1}^N  x^i H_i)  \
\exp ( \sum_{\Phi_+} v^\a E_{-\a }) \ .   $$
 $\Phi_+$ and $\Delta$ are the sets of the positive and simple roots of a
complex algebra with the Cartan-Weyl  basis consisting of the  step operators
$E_\a, \ E_{-\a}  , \ \a \in \Phi_+ $  and  $N(=$rank$ G)$  Cartan  subalgebra
generators $ H_\a = \a_i H_i , \ \a\in \Delta$.

The $four$ dimensional  ($D= 2+N=4$)
models
are obtained  for each  of the rank 2 maximally non-compact groups:
  $SL(3,R), \ SO(2,3)=Sp(4,R)$, $SO(2,2)=SL(2,R)\oplus SL(2,R) $   and $G_2$.
In the rank 2 case the corresponding background  is parametrised by  a $2\times
2$ Cartan
matrix or by  two simple roots with components $\a_{1i}$ and $\a_{2i}$
 and  one  parameter $\ep =\ep_2/\ep_1$ with values $ \pm 1$,
\eqn\ffffw{ L = {1 \ov   {\rm e}^{  \a_1\cdot x } + \ep  {\rm e}^{  \a_2\cdot x
 }  }
 \del u \bd v + \del x_1 \bd x_1 +
\del x_2 \bd x_2} $$ + \a' {\cal R} [\p_0  + \ha (\a_1\cdot x  +  \a_2\cdot x)
- \ha  \ln   \bigl(  {\rm e}^{  \a_1\cdot x  }   + \ep  {\rm e}^{  \a_2\cdot x
} \bigr)]  \ . $$
In addition to the Poincare symmetry in the $u,v$ plane this model
is also invariant under a correlated constant shift of $x^i$
and $u$ (or $v$).  For example, in the case of $SL(3,R)$
$\a_1=(\sqrt 2 , 0), \  \a_2=(- {1\ov \sqrt 2} , {3\ov \sqrt 2}) $.
For $\ep=1$ the curvature is non-singular in the coordinate patch used in
\ffff,
but there is also a horizon so one should first consider the geodesic
completion.
It is likely that there is still no curvature singularity, as in the case of
the $SL(2,R)$ group space.

The presence of the two chiral currents implies \klts\ that the
classical equations of these \sms (i.e. of  the string propagation in these
backgrounds)  reduce to the Toda equation for $x^i$
\eqn\tod{  \del\bd x_i  + \ha \chi \del_i F\inv =0 \ , \ \ \  F \bd v = \n
(\bar z)  \
, \  \ F \del u = \m (z)
\ , \  \  \chi \equiv
\n (\bar z) \m (z) \  . }
$\chi$ can be made constant by a coordinate transformation. Then
the  solutions  (including the solutions of the constraints)
can be  expressed in terms of the Toda model solutions.

Let us mention also that there exists another example of  \FM which
corresponds to a gauged WZW model:
the general  \FM in $D=3$. Solving \fff\ in $D=3$
one finds that $F\inv = a + m e^{bx}$.
As shown in \horts, this model can be obtained from a special
$SL(2,R)\times R /R$ gauged WZW model and, at the same time, is the extremal
limit
of the charged black string \hoho\  solution.
The case of $a=0$ corresponds to the ungauged $SL(2,R)$ WZW model
(with the action written in the Gauss decomposition parametrisation).

\subsec{ $D=4$  \FM with 2-dimensional euclidean black hole as the transverse
part}
It is possible to find another exact $D=4$ \FM solution by taking the
transverse
two dimensional theory to be non-trivial, i.e. represented by the $D=2$
euclidean black hole model. This solution is related to the \KM
with $K$ given by  \taa\  by the  $u$-duality transformation   \horts\
\eqn\fmo{ ds^2  = (a +  m\ \ln \th bx)^{-1} dudv  + dx^2 +
 { {\rm tanh}^2 bx }\ d{\theta}^2 \ ,   } $$  \ \ B_{uv} = \ha (a+ m\ \ln \th
bx)^{-1} \ , \ \ \   \p = \p_0  -  {\ln \ch }  bx + \ha \ \ln\ F \ . $$
This   background   can be viewed as a generalisation of  the  fundamental
string  \fss\  in four dimensions, $F\inv = 1 - M \ \ln\  r$. In addition to
the usual singularity at $r=0$ there is another singularity outside
the string at nonzero $r$. The solution \fmo\ has the same singularity at  the
origin (and hence can be viewed
as the field outside a fundamental string) but is regular elsewhere
and even asymptotically flat.
The original  fundamental string solution  can be recovered by
taking the limit $b \ra 0$ which is consistent
since the central charge condition is now imposed
 only at the level of the full $D=4$ solution.

\newsec{Further generalisations}
Let us consider the following generalisation of both the \KM and \FM
with flat transverse space
(cf. \stre,\mof)
\eqn\stree{   L =   F(x) \del u \bd v +
  K(x)\del u \bd u  +  \tilde A_i(x) \del u  \bd  x^i + \del x^i \bd x^i    +
\a'  {\cal R}  \p (x)\  ,  }
or ($\tilde A_i=FA_i$)
\eqn\stree{   L =   F(x) \del u [\bd v + A_i(x)   \bd  x^i]
+    K(x)\del u \bd u  +   \del x^i \bd x^i    +   \a'  {\cal R}  \p (x)\  .  }
This model has one null Killing vector, is invariant under the gauge
transformations of $A_i$ combined with a shift of $v$ and is `self-dual'.
In fact, it transforms into itself under the duality transformation  along a
direction in the $(u,v)$-plane
\eqn\dudu{
 F'= {F\ov K + qF}\ , \ \ \  K'= {1\ov K + qF} \ , \ \ \ A_i'= {A_i} \ , \ \
q=\const \ .  }
Just as for the \FM it is possible to prove \hrts\ that there exists a scheme
in which this model is conformally invariant to all orders provided
the one-loop conformal invariance conditions  are satisfied.
The latter are given by ($F_{ij}=\del_i A_j - \del_j A_i$)
\eqn\eqqq{   \del^2 F\inv=0 \ , \ \ \  \del^2 (KF\inv) =0 \ , \ \ \   \p=  \p_0
+ \ha  \ln F(x)\ , \ \ \  \del_i F^{ij}=0 \ .    }
There exists also  a straightforward generalisation to the case when $K$
depends on $u$.
As a results, one is able to find many new interesting
exact  solutions \hrts,   in particular,  various
generalisations of the fundamental string  solution
(some of them were already obtained as the leading-order solutions  in
\refs{\garf,\sen,\wald}).
In the case of solutions with the  spherical symmetry in the  transverse space
the solution for $K$ is
 $K=k + c F$ so that one can
redefine $v$ to make  $K=k=\const$.

\newsec{Concluding remarks}
We have discussed several new classes of exact classical
solutions in bosonic string theory.
They  can be embedded into the closed superstring theory
and thus also in the heterotic string theory (introducing appropriate
gauge field background equal to the spin connection).
All of the \KM and \FM solutions and their generalisations
 have  at least one null Killing vector.
This indicates  that backgrounds with null Killing vectors play special
role in string theory.
Though we have given several examples  when
such backgrounds can be derived from a gauged WZW model and thus  correspond to
a coset
CFT,
the question about  CFT interpretation of general \KM and \FM -type solutions
remains open. The fact that   the  exact expressions for these
backgrounds  are known explicitly certainly   suggests
  that the underlying CFT's should  exist and have well-defined properties.

This may not, however, apply to the fundamental string solution.
A peculiarity of the FS solution is that it admits two possible
interpretations. It can be  considered  as a vacuum solution  (an extremal
limit of a general class of charged string solutions \horstr) of the effective
equations  which is valid only outside the core $r=0$.
Alternatively, it can be  derived (as, in fact,  was  done in the original
approach of \gibb) as a  solution corresponding to the
combined action (we separate the constant part of the dilaton field $g_0=e^{
\p_0}$)
\eqn\edd{\hat S= {1\ov g^2_{0}} S_{eff} (\vp) + I_{str} (x; \vp)\ , \ \ \ \
I_{str}=I_0(x)  + V_i \vp^i \ , }
containing both the effective action for the background  fields  $\vp^i=
(G,B,\p)$ (condensates of massless string modes)  and the action
$I_{str} $
of a source string interacting with the background.  The source action
 leads to  the $\d (r)$-term
in the Laplace equation for $F\inv$.
Though such mixture of actions looks strange from the point of view of
perturbative  string theory,
 $\hat S$ can in fact be considered as  describing a non-perturbative
 `thin handle' (or `wormhole') approximation to quantum string partition
function  \tsmac:
\eqn\ook{ Z= \sum_{n=0}^\infty g_0^{2n-2} \int[d m]_n \int_{M_n} [dx]\   \exp
(-I_0 [x]) }
$$ \approx \int [d\vp] \int_{M_0} [dx]\   \exp [-  {1\ov g^2_{0}} S_{eff} (\vp)
- I_{str} (x; \vp)]
+ ... \    $$
Here $m$ denote  moduli, $M_n$ are 2-surfaces of genus $n$ and dots
indicate contributions of other parts of moduli spaces.
Extrema of $\hat S$ can thus be viewed as  some   non-perturbative
solitonic  solutions in string theory.  This raises the question if they can
actually be described by a conformal field theory since  we only know  that
solutions of the
tree-level effective equations  $\d S_{eff}/\d \vp=0$ correspond to  CFT's.

Given that  the effective string  field equations contain terms of all orders
in $\a'$
their general solution can be represented as ($\vp$ stands for a set
of `massless' fields): $ \vp=\vp_0 + \a' \vp_1 + \a'^2 \vp_2 + ... $.
Here $\vp_0$ is the leading-order  solution while
$\vp_n$ are, in general,
non-local (involving inverse Laplacians)  functionals of $\vp_0$.
It may happen that for some special $\vp_0$ the higher order corrections
are $local, covariant$  functionals of $\vp_0$. In that case one is able  to
change the scheme (i.e. to make a local covariant field redefinition) so that
in the new scheme  $\vp_0$ is actually  an exact solution.
It may be a bit disappointing to learn that all of the presently explicitly
known exact string solutions  are of that type, i.e. we do not yet know an
example
of a string solution  with truly non-trivial $\a'$-dependence (the one  which
cannot be absorbed into a local field redefinition).
It may happen that for  some leading-order solutions (e.g. Schwarzschild)
which can  of course  be deformed order by order in $\a'$ to make them satisfy
the
full string equations there is no $regular$ (satisfying standard axioms)   CFT
which corresponds to the
resulting $\a'$-series.\foot{Superstring  solutions  corresponding to \sms with
$n=2$ world sheet supersymmetry are examples of the situation where this does
not happen.
The corresponding $\beta$-function equations contain non-trivial
$\a'^3$- and higher- order corrections  \gris\ which cannot be redefined away
by a local redefinition of the $metric$ since otherwise one would  eliminate
the $R^4$-
term in the effective action  \grwi\ and thus change the string $S$-matrix.
The fact that there exists a local redefinition of the Kahler potential
\nemsen\   seems to
indicate that the $n=2$ case is special compared to the general  $n=1$.
In fact, here there are good reasons to expect
that the resulting $\a'$-deformed  solution  corresponds to a regular $n=2$
superconformal theory \gepner.}

\newsec{Acknowledgements}
I would  like to  thank
 G. Horowitz, R. Kallosh, C. Klim\v c\'\i k  and  K. Sfetsos for useful
discussions
and collaborations.
I  acknowledge also   the support of PPARC.

\vfill\eject
\listrefs
\end